\documentclass[10pt,letterpaper]{article}
\usepackage[top=0.85in,left=2.75in,footskip=0.75in]{geometry}
\usepackage[noend]{algpseudocode}
\usepackage{algorithmicx,algorithm}
\usepackage{bm}
\usepackage{multirow}
\usepackage{amsmath,amssymb}
\usepackage{changepage}
\usepackage{textcomp,marvosym}
\usepackage{cite}
\usepackage{nameref,hyperref}
\usepackage[right]{lineno}
\usepackage[nopatch=eqnum]{microtype}
\DisableLigatures[f]{encoding = *, family = * }
\usepackage[table]{xcolor}
\usepackage{array}
\newcolumntype{+}{!{\vrule width 2pt}}
\newlength\savedwidth

\raggedright
\usepackage[aboveskip=1pt,labelfont=bf,labelsep=period,justification=raggedright,singlelinecheck=off]{caption}

\makeatletter
\renewcommand{\@biblabel}[1]{\quad#1.}
\makeatother

\usepackage{lastpage,fancyhdr,graphicx}
\usepackage{epstopdf}
\fancyheadoffset[L]{2.25in}
\fancyfootoffset[L]{2.25in}

\begin{document}
\vspace*{0.2in}

\begin{flushleft}
{\Large
\textbf\newline{Low-dose CT reconstruction using dataset-free learning} 
}
\newline
\\
Feng~Wang\textsuperscript{*},
Renfang~Wang\textsuperscript{},
Hong~Qiu\textsuperscript{}
\\
\bigskip
\textbf{} College of Big Data and Software Engineering, Zhejiang Wanli University, Ningbo, Zhejiang, China
\\
\bigskip

* wangf\_721@zju.edu.cn

\end{flushleft}

\section*{Abstract}
Low-Dose computer tomography (LDCT) is an ideal alternative to reduce radiation risk in clinical applications. Although supervised-deep-learning-based reconstruction methods have demonstrated superior performance compared to conventional model-driven reconstruction algorithms, they require collecting massive pairs of low-dose and norm-dose CT images for neural network training, which limits their practical application in LDCT imaging. In this paper, we propose an unsupervised and training data-free learning reconstruction method for LDCT imaging that avoids the requirement for training data. The proposed method is a post-processing technique that aims to enhance the initial low-quality reconstruction results, and it reconstructs the high-quality images by neural work training that minimizes the $\ell_1$-norm distance between the CT measurements and their corresponding simulated sinogram data, as well as the total variation (TV) value of the reconstructed image. Moreover, the proposed method does not require to set the weights for both the data fidelity term and the plenty term. Experimental results on the AAPM challenge data and LoDoPab-CT data demonstrate that the proposed method is able to effectively suppress the noise and preserve the tiny structures. Also, these results demonstrate the rapid convergence and low computational cost of the proposed method. The source code is available at \url{https://github.com/linfengyu77/IRLDCT}.


\section*{Introduction}
X-ray computed tomography (CT) is an essential imaging modality for clinical purposes, as it provides high-resolution images of the internal structure of the human body. However, X-ray radiation is known to be harmful to healthy tissues. In some major clinical tasks, a single CT scan can expose patients to radiation doses of up to 43 mSv \cite{smith-bindman_radiation_2009}, which may increase the risk of cancer. Consequently, reducing radiation dose while obtaining high-resolution images has become a significant area of research in CT scanning. 

Currently, there are two primary strategies for reducing CT radiation dose: (1) decreasing the number of projection views and (2) lowering the X-ray tube current. This approach is commonly referred to as LDCT. LDCT algorithms can be broadly categorized into four groups: sinogram domain filtering, iterative reconstruction, and deep learning-based reconstruction.

Sinogram domain filtering methods exploit the distinct distributions of desired signals and noise in the sinogram domain to reconstruct CT images. This technique involves filtering out components corresponding to artifacts or noise in the sinogram domain and then inverting the filtered sinogram data into the image domain using analytic algorithms. Numerous analytic filtering methods have been proposed based on the distribution of noise. For instance, filtered back projection (FBP) is a classical reconstruction method for CT images that performs high-pass filtering in the sinogram domain before back-projection. Sinogram domain filtering can produce high-quality CT images when the noise distribution is accurately characterized. However, determining this distribution can be challenging, particularly since artifacts or noise often correlate with image structures.

Compared with sinogram domain filtering methods, iterative reconstruction approaches are more flexible and stable. Iterative reconstruction approaches can be further divided into hybrid iterative reconstruction methods and model-based iterative reconstruction methods. Hybrid iterative reconstruction method produces an image by adjusting the statistical characters of the sinogram domain and the image domain. Model-based iterative reconstruction method utilizes the process of alternative performing the forward-projection (i.e., sinogram data generation) and back-projection (i.e., CT image reconstruction) to achieve iterative filtering in the sinogram domain and the image domain. Furthermore, the cost function of model-based iterative reconstruction method usually consists of a fidelity term with the noise model in the sinogram domain and a regularization term with the prior model in the image domain. The regularization term plays a vital role in reconstruction, and many regularizations have been proposed, such as total variation (TV)\cite{sidky_image_2008,kim_sparse-view_2015}, low-rank \cite{cai_cine_2014}, non-local means (NLM) \cite{ma_iterative_2012, zhang_spectral_2016}, and dictionary learning \cite{qiong_xu_low-dose_2012}. The model-based iterative reconstruction method usually has better performance than hybrid iterative reconstruction method, but it is also computationally expensive. Additionally, model-based iterative reconstruction method requires manually designing the proper regularization and choosing the weight to obtain satisfactory reconstruction results.

\begin{figure*}[!t]
\centering
\includegraphics[width=\textwidth]{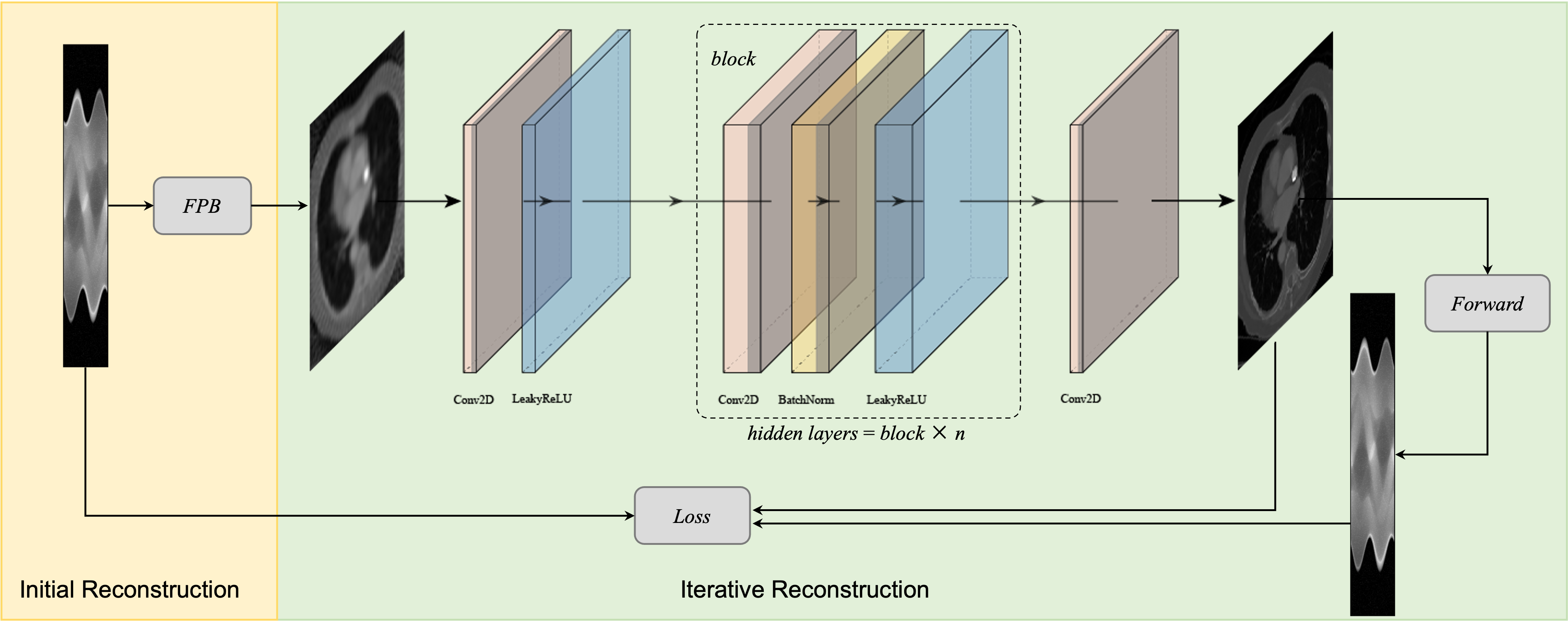}
\caption{Schematic diagram of the proposed method.}
\label{dig}
\end{figure*}

In recent years, deep learning techniques have been widely employed in LDCT reconstruction, and they have demonstrated better performance than conventional LDCT reconstruction methods. Deep learning-based LDCT reconstruction methods can be categorized into four groups: sinogram domain processing (SDP), image domain processing (IDP), dual-domain processing (DDP), sinogram-image direct mapping (SIDP), and model-based deep learning (MBDL).

The SDP reconstruction algorithm aims to use a pre-trained neural network to inpaint the LDCT measurements into sinogram data that is very close to normal-dose CT (NDCT) measurements. For instance, \cite{yuan_low-dose_2019} proposed a sinogram domain denoising approach using a convolutional neural network (CNN) with a filter loss function. Compared with image domain denoising methods, these approaches can easily estimate the noise level in the projection. Reference \cite{liu_sparse-sampling_2020} proposed a sinogram data interpolation method by leveraging a conditional adversarial network (GAN). Although sinogram domain processing can correct errors in the sinogram domain, errors produced by the shortcomings of conventional methods can still negatively affect the final reconstructions.

In contrast to the SDP algorithm, IDP produces high-quality CT images by using a neural network to denoise the initial reconstructed images with artifacts. Most deep learning methods employ IDP to improve the quality of reconstructed images obtained using existing methods such as FBP \cite{jin_deep_2017,chen_low-dose_2017}. Reference \cite{hasan_hybrid-collaborative_2021} introduced a collaborative technique to train multiple Noise2Noise \cite{lehtinen_noise2noise_2018} generators simultaneously and learn the image representation from LDCT images. Reference \cite{hendriksen_noise2inverse_2020} proposed Noise2Self that does not require any additional clean or noisy data. IDP is more straightforward compared to the SDP algorithm. Reference \cite{evangelista_rising_2023} proposed to a framework for sparse-view tomographic image reconstruction combining an early-stopped rapid iterative solver with a subsequent pre-trained neural network to complete the missing iterations of rapid iterative solver. One main disadvantage of IDP is that it is difficult to recover information lost from the initial reconstructed images, which serve as inputs to the neural network.

DDP is a method that combines SDP and IDP. It leverages the advantages of both SDP and IDP to achieve higher-quality images compared to single-domain processing reconstruction methods. Reference \cite{kang_deep_2017} combined a deep convolutional neural network (CNN) with directional wavelet transform to extract the directional component of artifacts in low-dose CT images and exploit intra- and inter-band correlations. Reference \cite{zheng_dual-domain_2020} proposed a deep learning-based function optimization method for LDCT imaging, which incorporated the Radon inverse operator and disentangled each slice. To address of the limitation of acquiring independent noisy reference image of Noise2Noise \cite{lehtinen_noise2noise_2018}, \cite{yuan_half2half_2020} proposed a method to generate both training inputs and training labels from the existing CT scans for count-domain and image-domain, which does not require any additional high-dose CT images or repeated scans. Although DDP can achieve good inversion results, it requires a larger training dataset due to its two training procedures: sinogram domain and image domain.

SIDP is an end-to-end reconstruction algorithm that directly transforms sinogram data into CT images. This method has the lowest complexity as it only requires training a neural network without extra processing such as sinogram data correction and inversion. For example, \cite{zhu_image_2018} presented a unified framework for image reconstruction called Automated Transform by Manifold Approximation (AUTOMAP), which directly converts sinogram data into CT images. Reference \cite{kandarpa_dug-recon_2021} proposed a direct reconstruction framework exclusively using deep learning architectures, which consists of denoising, reconstruction, and super resolution (SR). SIDP is a highly efficient reconstruction method but demands massive memory as the entire sinogram data needs to be fed into the neural network.

MBDL, also known as optimization unrolling scheme or plug-and-play, is an effective approach that replaces the parameters or regularization of conventional iterative schemes with learnable/pre-trained neural networks. Reference \cite{wu_computationally_2019} unrolled the proximal gradient descent algorithm for iterative image reconstruction to finite iterations and replaced terms related to the penalty function with trainable CNN to reduce memory requirements and training time. Reference \cite{cheng_learned_2020} incorporated the benefits from analytical reconstruction methods, iterative reconstruction methods, and DNNs. They unrolled proximal forward-backward splitting into iterative reconstruction updates of CT data fidelity and DNN regularization with residual learning. Reference \cite{ye_unified_2021} developed a unified reconstruction framework combining supervised and unsupervised learning, and physics and statistical models to enhance the accuracy and resolution of LDCT reconstruction images. By leveraging the advantages of deep learning and conventional methods, MBDL offers better interpretability than data-driven deep learning.

\begin{figure*}[]
\centering
\includegraphics[width=\textwidth]{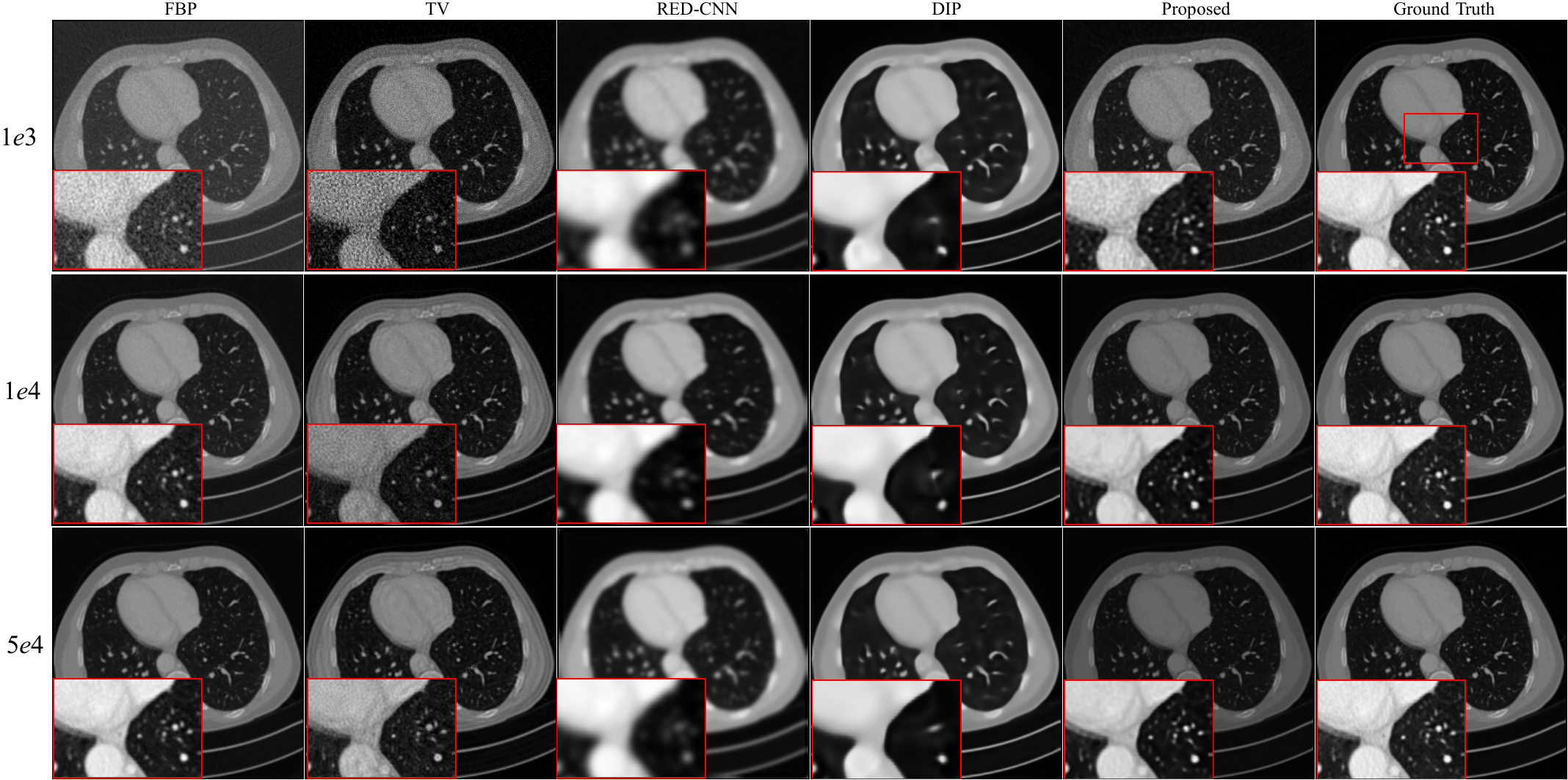}
\caption{Reconstruction results of case AAPM-1 at different dose levels by different methods. Zoomed parts over the region of interest (ROI) marked by the red box in the ground-truth image.}
\label{aapm1_reco}
\end{figure*}

\begin{figure*}[]
\centering
\includegraphics[width=\textwidth]{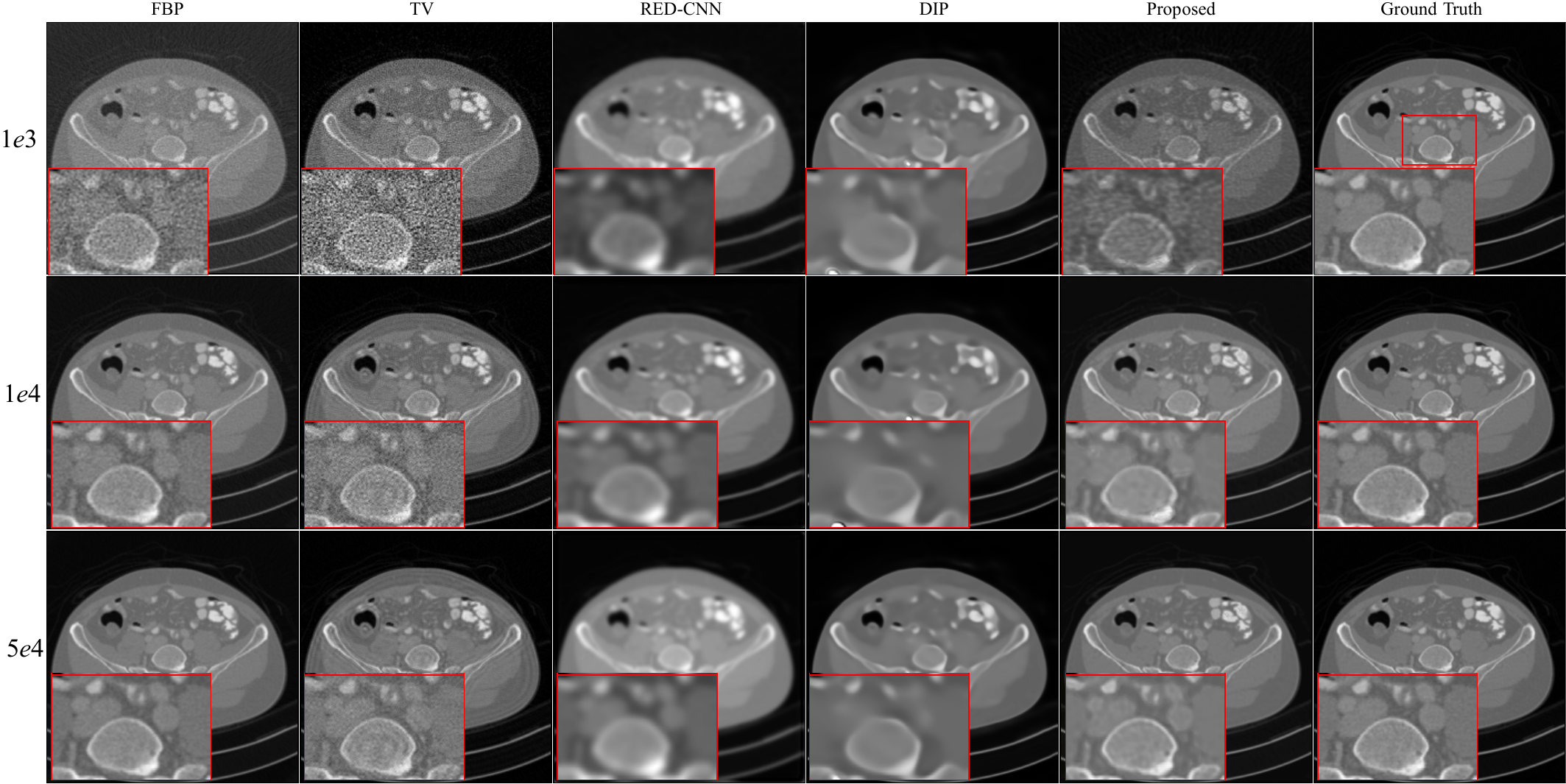}
\caption{Reconstruction results of case AAPM-2 at different dose levels by different methods. Zoomed ROI images from the ground-truth image.}
\label{aapm2_reco}
\end{figure*}

\begin{figure*}[]
\centering
\includegraphics[width=\textwidth]{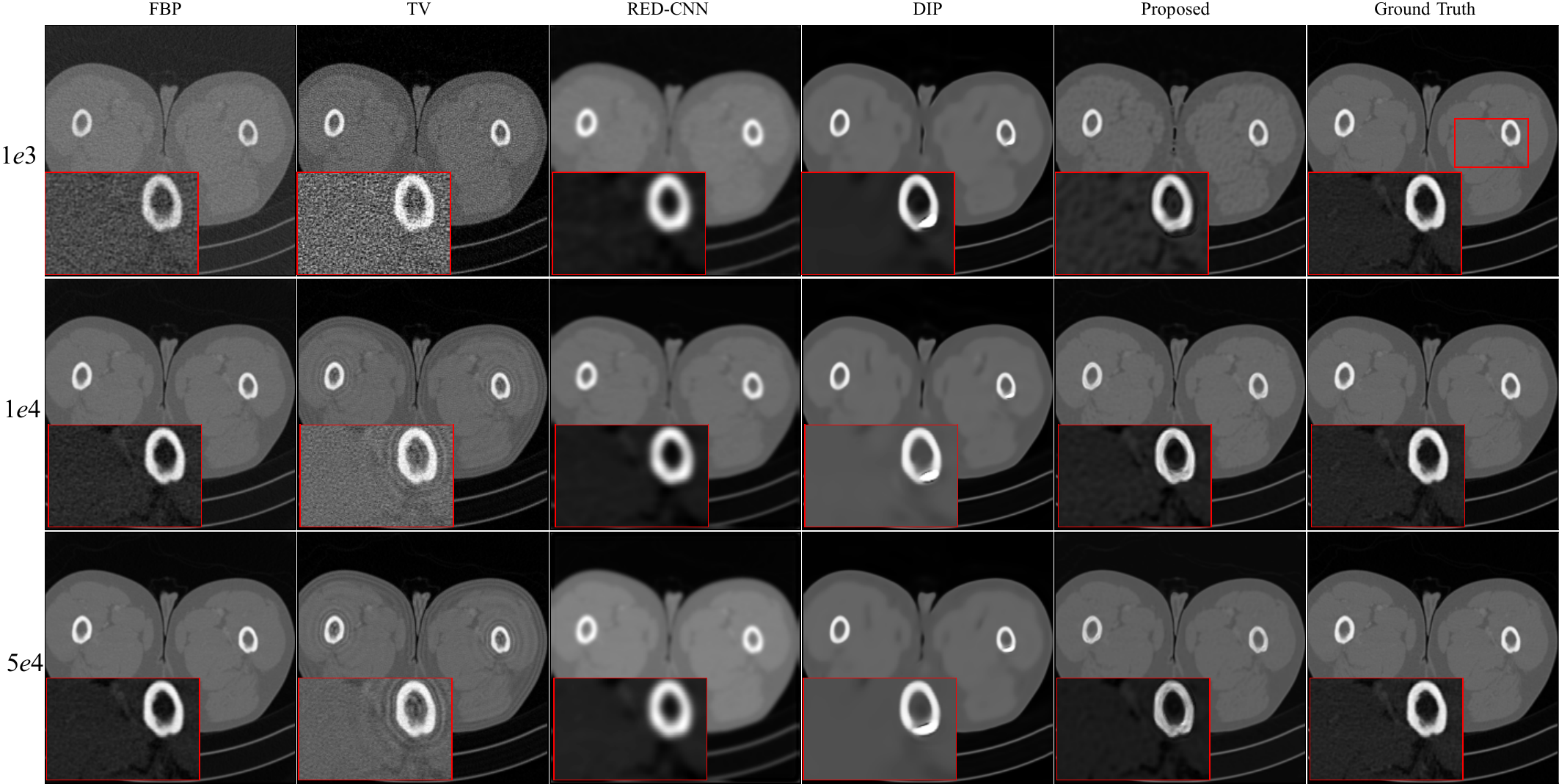}
\caption{Reconstruction results of case AAPM-3 at different dose levels by different methods. Zoomed ROI images from the ground-truth image.}
\label{aapm3_reco}
\end{figure*}

Recently, training dataset-free method have drawn much attention in LDCT imaging, which does not need to pre-train a neural network and works on a single image by utilizing the consistency between the CT measurements and sinogram data modeled on the reconstructed image. For instance, the deep image prior (DIP) \cite{ulyanov_deep_2020}, originally proposed for natural image denoising by using early stopping to fit the noisy image, has been widely exploited in medical imaging \cite{gong_pet_2019,yokota_dynamic_2019,yoo_time-dependent_2021}. Also, DIP treats noise as i.i.d random noise rather than artifacts correlated to the entries of CT images. Reference \cite{ding_dataset-free_2022} proposed an dataset-free reconstruction method based on Bayesian inference, which takes the $\mathcal{J}$-invariant transform of the FBP reconstructed image as the initial value. This method can reconstruct high-quality images from measurements; however, its reconstruction time is significantly higher than that of its competitors.

In this paper, we propose an iterative LDCT reconstruction method that ultizes neural network to improve the CT images reconstructed by FBP method without training data. During the iterative LDCT reconstruction, we minimize the loss, which consists of two components: the $\ell_1$-norm distance between the CT measurements and the sinogram data modeled on the post-processed image, and the TV value of the post-processed image. We achieve this by training a neural network. The proposed method does not require collecting any training data and balancing the contribution of data fidelity and TV regularization in the loss. Once the network training is complete, the high-quality reconstructed results will be output immediately.

The rest of the paper is organized as follows: Methodology section describes how to build and solve the optimization problem. Experimental Results section presents the experimental setup and results using the 2016 Low-dose CT Grand Challenge data and the LoDoPaB-CT data \cite{leuschner_lodopab-ct_2021}. Discussion and Conclusion section is the discussion and conclusion.

\section*{Methodology}
In this section, we introduce a proposed method for reconstructing LDCT from noisy measurements. This method utilizes a DNN to enhance the CT image reconstructed by the FBP method, without the need for training data.

\subsection*{Problem Setup}
The forward formulation of LDCT can be formulated as

\begin{equation}
    \bm{y}=\bm{A}\bm{x} + \bm{\epsilon},
    \label{eq1}
\end{equation}

\noindent where $\bm{y}$ represents the CT measurements, $\bm{A}$ is the projection matrix of CT imaging, $\bm{\epsilon}$ denotes the background contributions of scatter and electrical noise, and $\bm{x}$ represents the ground-truth CT image. Typically, we can solve the inverse problem of Eq.~\ref{eq2} by using the FBP method $\mathcal{F}$, 

\begin{equation}
    \bm{x}_{f}=\mathcal{F}\left(\bm{y}\right),
    \label{eq2}
\end{equation}

\noindent where $\bm{x}_{f}$ is the reconstructed image by FBP. However, due to the low source intensity of X-ray and/or the random noise, the quality of $\bm{x}_{fbp}$ is unsatisfactory, often suffering from noticeable streaky artifacts, random patterns, and low resolution.

Considering a DNN $\mathcal{NN}$ with parameters $\bm{\theta}$ that can enhance the image's quality by $\mathcal{NN}\left(\bm{x}_{f}; \bm{\theta}\right)$, which means that we can re-formulate Eq.~\ref{eq1} as

\begin{equation}
    \bm{y}=\bm{A}\mathcal{NN}\left(\bm{x}_{f}; \bm{\theta}\right) + \bm{\epsilon},
    \label{eq3}
\end{equation}

According to Bayes's rule, we can obtain the posterior density of $\mathcal{NN}\left(\bm{x}_{f}; \bm{\theta}\right)$ by

\begin{equation}
    p(\mathcal{NN}\left(\bm{x}_{f}; \bm{\theta}\right) \mid \bm{y})\propto p( \bm{y} \mid \mathcal{NN}\left(\bm{x}_{f}; \bm{\theta}\right))p(\bm{y}).
    \label{eq4}
\end{equation}

Supposing $p( \bm{y} \mid \mathcal{NN}\left(\bm{x}_{f}; \bm{\theta}\right))$ as Gaussian distribution,

\begin{equation}
    p( \bm{y} \mid \mathcal{NN}\left(\bm{x}_{f}; \bm{\theta}\right))= \mathcal{N}(\bm{A}\mathcal{NN}\left(\bm{x}_{f}; \bm{\theta}\right), \boldsymbol{\Sigma}_{\epsilon}).
    \label{eq5}
\end{equation}

\noindent where $\boldsymbol{\Sigma}_{\epsilon}$ represents the covariance of the noise. Furthermore, taking the logarithm on both sides of Eq.~\ref{eq5} then we obtain

\begin{equation}
    \begin{aligned}
    \ln p(\bm{y} \mid \mathcal{NN}\left(\bm{x}_{f}; \bm{\theta}\right)) &\propto \frac{1}{2}\left(\bm{y} - \bm{A}\mathcal{NN}\left(\bm{x}_{f}; \bm{\theta}\right)\right)^\mathrm{T}\boldsymbol{\Sigma}^{-1}_{\epsilon}\left(\bm{y} - \bm{A}\mathcal{NN}\left(\bm{x}_{f}; \bm{\theta}\right)\right).
    \end{aligned}
    \label{eq6}
\end{equation}

Taking the logarithm on both sides of Eq.~\ref{eq4} and substituting $\ln p(\bm{y} \mid \mathcal{NN}\left(\bm{x}_{f}; \bm{\theta}\right))$ with Eq.~\ref{eq6}, we obtain

\begin{equation}
    \begin{aligned}
    \ln p(\mathcal{NN}\left(\bm{x}_{f}; \bm{\theta}\right) \mid \bm{y}) &\propto \ln p(\mathcal{NN}\left(\bm{x}_{f}; \bm{\theta}\right) \mid \bm{y})p(\mathcal{NN}\left(\bm{x}_{f}; \bm{\theta}\right))\\
    &\propto \frac{1}{2}\left(\bm{y} - \bm{A}\mathcal{NN}\left(\bm{x}_{f}; \bm{\theta}\right)\right)^\mathrm{T}\boldsymbol{\Sigma}^{-1}_{\epsilon}\left(\bm{y} - \bm{A}\mathcal{NN}\left(\bm{x}_{f}; \bm{\theta}\right)\right) + \ln p\left(\bm{y}\right).
    \end{aligned}
    \label{eq7}
\end{equation}

Therefore, we can obtain the maximum a posterior (MAP) objective,

\begin{equation}
    \begin{aligned}
        \max \left\{ \ln p(\mathcal{NN}\left(\bm{x}_{f}; \bm{\theta}\right) \mid \bm{y}) \right\} &= \min \left\{ -\ln p(\mathcal{NN}\left(\bm{x}_{f}; \bm{\theta}\right) \mid \bm{y}) \right\}\\
        &\propto \min \left\{ \frac{1}{2}\left(\bm{y} - \bm{A}\mathcal{NN}\left(\bm{x}_{f}; \bm{\theta}\right)\right)^\mathrm{T}\boldsymbol{\Sigma}^{-1}_{\epsilon}\left(\bm{y} - \bm{A}\mathcal{NN}\left(\bm{x}_{f}; \bm{\theta}\right)\right)\right\}.
    \end{aligned}
    \label{eq8}
\end{equation}

\noindent Assuming the noise is Gaussian independent and identically distributed (iid), i.e., $\Sigma_{\epsilon}=\sigma_{\epsilon}^{2}\mathbf{I}$. Furthermore, considering to regularize $\mathcal{NN}\left(\bm{x}_{f}; \bm{\theta}\right)$ with $\mathcal{R}(\cdot)$, Eq.~\ref{eq8} can be further rewritten as

\begin{equation}
    \begin{aligned}
        \bm{x}^{\ast} = \underset{\bm{\theta}}{\arg \min}\left\{ \frac{1}{2\sigma^{2}_\epsilon}\left(\bm{y} - \bm{A}\mathcal{NN}\left(\bm{x}_{f}; \bm{\theta}\right)\right)\left(\bm{y} - \bm{A}\mathcal{NN}\left(\bm{x}_{f}; \bm{\theta}\right)\right)\right \} \\ \text{subject to}\quad \eta\mathcal{R}(\mathcal{NN}\left(\bm{x}_{f}; \bm{\theta}\right)),
    \end{aligned}
 \label{eq9}
\end{equation}

\noindent where $\eta$ denotes the weight. To linearize this problem, we reformulate Eq.~\ref{eq9} as

\begin{equation}
    \bm{x}^{\ast} = \underset{\bm{\theta}}{\arg \min}\left\| \bm{y} - \bm{A}\mathcal{NN}\left(\bm{x}_{f}; \bm{\theta}\right) \right \|^{2}_{2} + \eta\mathcal{R}\left(\mathcal{NN}\left(\bm{x}_{f}; \bm{\theta}\right)\right).
    \label{eq10}
\end{equation}

\noindent In fact, the artifacts in LDCT images are highly correlated to the entries of CT images rather than random noise, and the results inverted through $\ell_2$-norm loss tend to be over-smoothed, which is not beneficial for preserving the tiny structures and/or sharp edges. Hence, we propose to optimize $\bm{\theta}$ by minimizing the $\ell_1$-norm misfit,

\begin{equation}
   \bm{x}^{\ast}=\arg \min _{\bm{\theta}}\|\bm{y} - \bm{A}\mathcal{NN}\left(\bm{x}_{f}; \bm{\theta}\right)\|_1 + \eta\mathcal{R}\left(\mathcal{NN}\left(\bm{x}_{f}; \bm{\theta}\right)\right).
    \label{eq11}
\end{equation}

Furthermore, we add the TV term of the reconstructed CT image $\hat{\bm{x}}=\mathcal{NN}\left(\bm{x}_{f}; \bm{\theta}\right)$ into Eq.~\ref{eq11} as a smooth penalty to overcome the potential over-fitting induced by the noise in the CT measurements. Eq.~\ref{eq11} thus becomes

\begin{equation}
    \begin{array}{l}
        {\begin{aligned}
                \bm{x}^{\ast}=\arg \min _{\bm{\theta}}\left\{\|\boldsymbol{y}-\boldsymbol{A} \mathcal{NN}\left(\bm{x}_{f}; \bm{\theta}\right)\|_1 + \nabla(\mathcal{NN}\left(\bm{x}_{f}; \bm{\theta}\right))\right\},
        \end{aligned}}\\
        {\begin{aligned}
            \nabla(\hat{\bm{x}})=\frac{1}{NM}(\sum\limits_{i=1}^{N}\sum\limits_{j=1}^{M-1}\|\hat{\bm{x}}_{i,j+1}-\hat{\bm{x}}_{i,j}\|_1  +\sum\limits_{i=1}^{N-1}\sum\limits_{j=1}^{M}\|\hat{\bm{x}}_{i+1,j}-\hat{\bm{x}}_{i,j}\|_1),
        \end{aligned}
        }
    \end{array}
    \label{eq12}
\end{equation}

\noindent where $\hat{\bm{x}} \in \mathrm{R}^{N \times M}$. Eq.~\ref{eq12} can be solved by NN training, and we can derive $\boldsymbol{x}^{\ast}$ with the forward propagation of $\mathcal{NN}$ once $\bm{\theta}$ be optimized by

\begin{equation}
    \boldsymbol{x}^{\ast}=\mathcal{NN}(\bm{x}_{f}; \boldsymbol{\theta}^{\ast}).
    \label{eq13}
\end{equation}

\subsection*{Solving the MAP}
The proposed method can be considered as a kind of NN training-based reconstruction method, which optimizes the NN's parameters by minimizes the loss from both sinogram domain and image domain. The proposed LDCT reconstruction method can be divided into two steps: (1) Reconstructing the initial CT image: The initial CT image is reconstructed using the the FBP method. Although this initial CT image may contain many artifacts due to the low intensity of X-ray, FBP provides fundamental information about the internal structure of the human body, which is helpful for enhancing the reliability of the inversion result by NN. Moreover, FBP performs much faster than iterative reconstruction approaches such as compressive sensing; (2) Post-processing the initial reconstruction result: Once the initial CT image is achieved, it will be fed into a pre-defined NN and will be improved through the NN training. To achieve $\boldsymbol{\theta}^{\ast}$, we establish the loss function for NN training based on Eq.~\ref{eq12}, 

\begin{equation}
    \mathcal{L}(\bm{y},\bm{x}_{f})=\frac{1}{NM}\|\boldsymbol{y}-\boldsymbol{A} \mathcal{NN}(\bm{x}_{f}; \boldsymbol{\theta})\|_1 + \nabla(\mathcal{NN}(\bm{x}_{f}; \boldsymbol{\theta})),
    \label{eq14}
\end{equation}

\noindent and we use gradient descent-based optimization algorithms such as stochastic gradient descent to optimize $\boldsymbol{\theta}$ to minimize the loss function. It is worth noting that the proposed method does not require setting weights for both the data fidelity term and the regularization term, which significantly reduces the difficulty of manually setting the weights. In summary, the visual flowchart of the iterative reconstruction for LDCT is shown in Fig.~\ref{dig}. Algorithm~\ref{alg} further explains the construction algorithm in detail.

\begin{algorithm}
\caption{Iterative reconstruction for LDCT imaging}
\label{alg}
\begin{algorithmic}
  \Require number of iterations: $n$; CT measurements: $y$; FPB operator: $\mathcal{F}$; projection matrix: $\bm{A}$; learning rate: $\lambda$
   \Ensure optimal $\boldsymbol{\theta}^{\ast}$
    \State initial $i=1$, initial CT image $x_0=\mathcal{F}(y)$
    \While{$i <= n$}
          \State reconstruct the CT image $\mathcal{NN}(\bm{x}_0;\bm{\theta})$
          \State compute the loss $ \mathcal{L}(\bm{y})=\frac{1}{NM}\|\boldsymbol{y}-\boldsymbol{A} \mathcal{NN}(\bm{x}_0 ; \boldsymbol{\theta})\|_1 + \nabla(\mathcal{NN}(\bm{x}_0 ; \boldsymbol{\theta}))$
          \State \text{update} $\bm{\theta}^{\ast} \gets \bm{\theta} - \lambda\frac{\partial\mathcal{L}}{\partial\bm{\theta}}$
          \State $i \gets i+1$
    \EndWhile
    \State return $\boldsymbol{x}^{\ast}= \mathcal{NN}(\bm{x}_0 ; \boldsymbol{\theta}^{\ast})$
\end{algorithmic}
\end{algorithm}

\subsection*{NN Architecture}
To enhance the quality of the initial CT image, we have designed a DNN with a straightforward structure. As depicted in Fig.~\ref{dig}, the network primarily consists of 2-D convolution, batch normalization, and LeakyReLU layers. The first layer is a convolution layer, followed by a LeakyReLU layer and a block composed of convolution, batch normalization (BN), and LeakyReLU. The convolution layers are employed for feature extraction, the BN layers for enhancing the stability of network training, and the LeakyReLU layers to ensure non-linearity throughout the network. LeakyReLU is defined as,

\begin{equation}
    \text{LeakyReLU}(x)=\max(0,x)+\phi \ast \min(0,x).
    \label{eq15}
\end{equation}

\noindent In the subsequent experimental test, we set the value of $\phi$ to 0.01.

\begin{figure*}[]
\centering
\includegraphics[width=\textwidth]{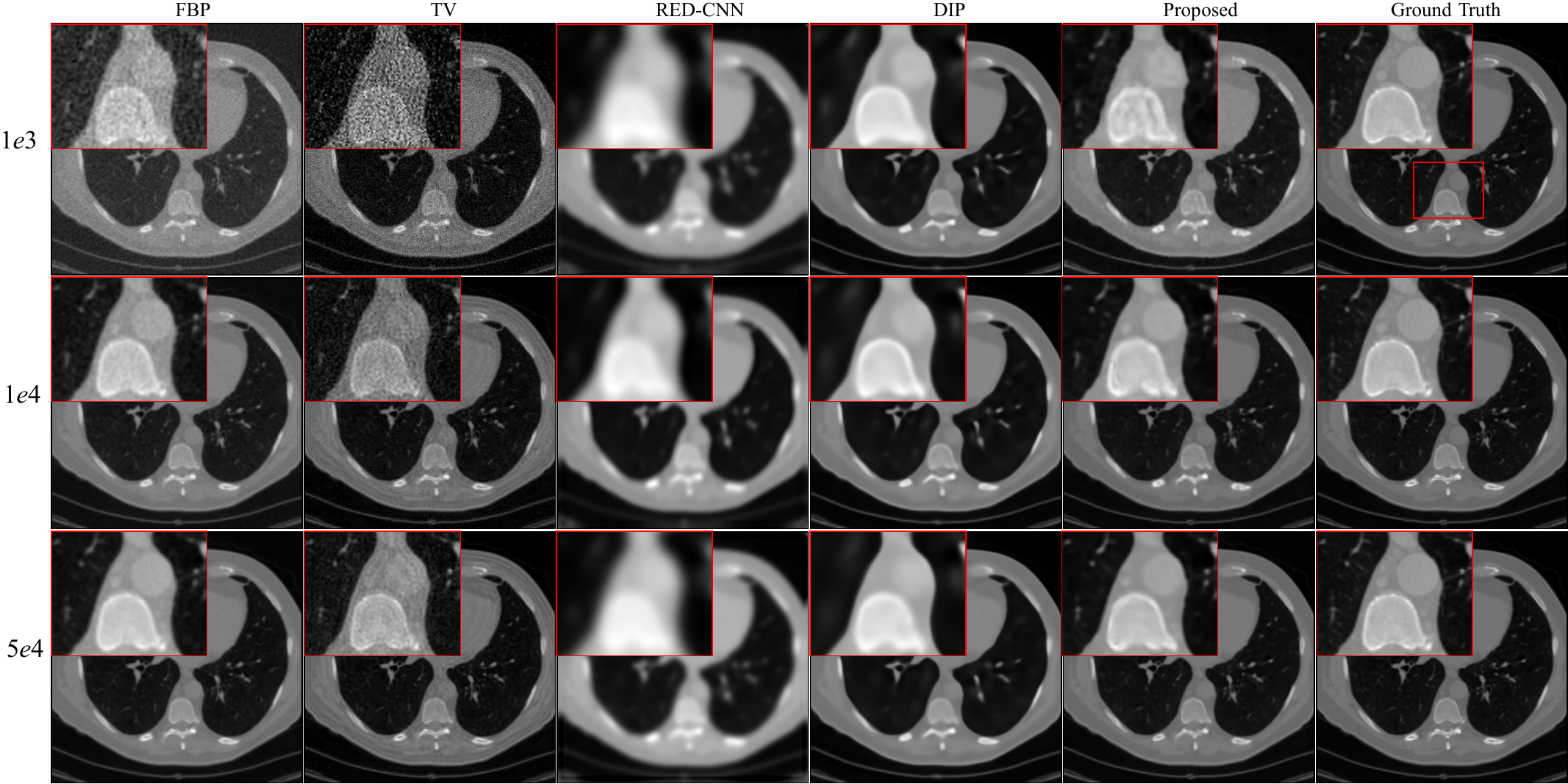}
\caption{Reconstruction results of LoDoPab-CT data at different dose levels by different methods. Zoomed ROI images from the ground-truth image.}
\label{lodopab_reco}
\end{figure*}

\begin{figure*}[]
\centering
\includegraphics[width=\textwidth]{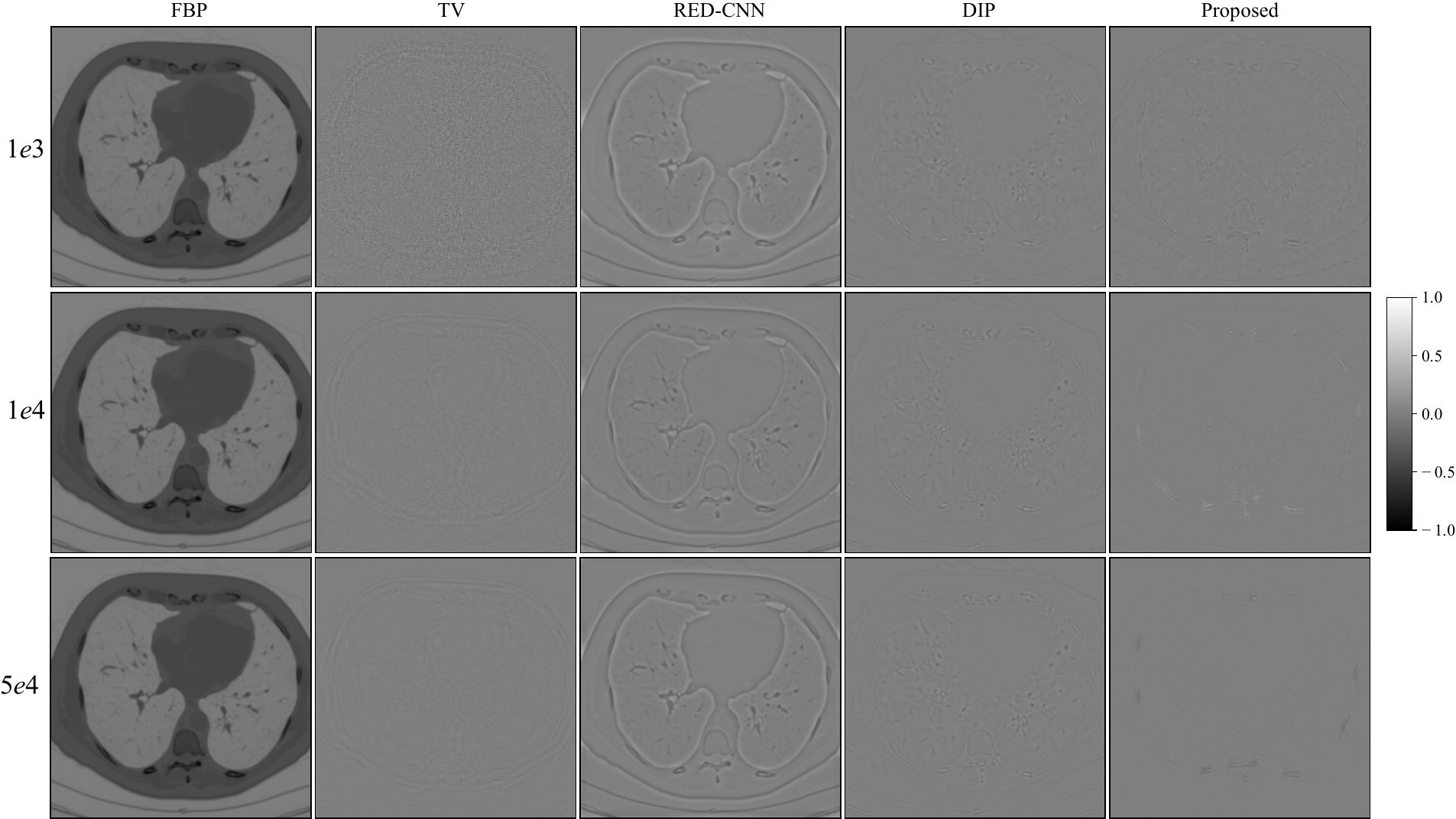}
\caption{Reconstruction errors of LoDoPab-CT data at different dose levels by different methods.}
\label{lodopab_errors}
\end{figure*}

\section*{Experimental Results}
In this section, we evaluate the performance of the proposed method by comparing it with four representative methods: FBP, TV (post-processing and unsupervised method), DIP (unsupervised and data-free method), and RED-CNN (post-processing and supervised model).

\subsection*{Parameter Setting}
The FBP, TV and DIP reconstruction are implemented by using Deep Inversion Validation Library (\url{https://github.com/jleuschn/dival}) and Operator Discretization Library (\url{https://github.com/odlgroup/odl}). For TV and DIP reconstruction, we use the parameters recommend by \cite{baguer_computed_2020} (\url{https://github.com/oterobaguer/dip-ct-benchmark}). The NN training related tasks are all implemented on the PyTorch platform \cite{paszke_pytorch_2019}.  

For TV reconstruction, the weight for $\ell_1$-norm term is set to $2.15 \times 10^{-7}$, and the number of iterations is set to 200, and we utilize the Douglas-Rachford Primal-Dual method as the solver. In addition, the initial reconstruction results for the TV method are obtained by the FBP method, and the parameters for the FBP method are the same as those for the proposed method, which means that both the TV method and the proposed method have the same input for the neural network. For DIP reconstruction, we use a learning rate of 0.0005, 6 scales, 1000 iterations for AAPM challenge data and 2000 iterations for LoDoPaB-CT data, and 128 channels for the U-Net at every scale.  We adopt mean square error (MSE) as the loss function for both TV and DIP reconstruction. In the proposed method, we set the iterations to 2000, and save the result with the highest peak signal-to-noise ratio (PSNR). For all reconstruction methods, the filter and frequency scaling of FBP reconstruction are set to Hann and 0.8, respectively.

For RED-CNN training, we use the AAPM Challenge Data as the training dataset. We train the RED-CNN using full-dose CT scans from nine patients, reserving one patient (L067) for evaluation. In the training data generation process, we use a patch size of 64. The batch size for RED-CNN training is set to 32, the number of training epochs is 100, the loss function is MSE loss, and we use Adam optimizer with a learning rate of $10^{-5}$. We train three models for different low-dose levels by using pairs of FBP reconstructions of low-dose simulations and corresponding full-dose CT images.

There are 30 convolution layers in our NN, the size of filter kernels of the first convolution layer is $64\times1\times3\times3$, where the format is $\text{number of filters} \times \text{number of channels} \times \text{width} \times \text{height}$. From the second to the penultimate convolution layer, we set the size of all filter kernels to $64\times 64\times 3 \times 3$. For the last convolution layer, the size of filter kernels is set to $1\times 64 \times 3 \times 3$. We minimize the loss defined by Eq.~\ref{eq8} by using the AdamW method with learning rate of $10^{-3}$.

\subsection*{Data Specification}
To evaluate the effectiveness of the proposed method, we test its performance on on two datasets: AAPM challenge data and LoDoPaB-CT data \cite{leuschner_lodopab-ct_2021}. The AAPM challenge data consists of reconstructed simulated data from human abdomen CT scans provided by Mayo Clinic for the AAPM Low Dose CT Grand Challenge (\url{https://www.aapm.org/GrandChallenge/LowDoseCT/}). We use 1-mm slice thickness reconstructions with  dimensions of 512 px $\times$ 512 px for RED-CNN training and performance comparison. The CT images form LoDoPaB-CT data are sampled from AAPM challenge data and have been cropped to dimensions of 362 px $\times$ 362 px. Additionally, these images have been subjected to dequantization noise uniformly distributed in [0,1] for each pixel.

For sinogram data simulation, we construct a 2-D fan-beam geometry with 1000 angles, 1000 pixels, source to axis distance 500 mm, and axis to detector distance 500 mm \cite{he_iterative_2022}. The LDCT image are simulated by adding Poisson noise with $I_i=[1e3, 1e4, 5e4]$ following the Poisson distribution according to the process of photon generation, attenuation, and detection, which can be expressed as,

\begin{equation}
    \bm{y}_i \sim \text{Poisson}\left\{I_i e^{-[\bm{Ax}]_i}+\bm{\sigma}_i\right\}, i=1, \ldots, m,
    \label{eq16}
\end{equation}

\noindent where $I_{i}$ denotes the source intensity of the \emph{i}-th X-ray, $\bm{y}_i$ represents the CT measurements produced by the \emph{i}-th X-ray, $\bm{A}$ is the projection matrix of CT imaging, $\bm{\sigma}_i$ denotes the background contributions of scatter and electrical noise, and $\bm{x}$ represents the full-dose CT image. Additionally, the full-dose CT images $x$ are normalized before sinogram simulation by  

\begin{equation}
    x=\frac{x-\min (x)}{\max(x)-\min(x)}.
    \label{eq17}
\end{equation}

\subsection*{Quantitative Indices}
We adopt two quantitative indices, PSNR and structural similarity index (SSIM), to quantify the quality of the reconstructed CT images. The PSNR expresses the ratio between the maximum possible power of a signal and the power of corrupting noise, which is measured by the mean squared error (MSE),

\begin{equation}
\begin{array}{l}
     {\begin{aligned}
         \text{PSNR}(\widetilde{x}, x)=10 \log _{10}\left(\frac{\max (x)^2}{\text{MSE}(\widetilde{x}, x)}\right),
     \end{aligned}}  \\
     {\begin{aligned}
          \text{MSE}(\widetilde{x}, x)=\frac{1}{n} \sum_{i=1}^n\left\|\widetilde{x}_i-x_i\right\|^2,
     \end{aligned}} 
\end{array}
    \label{eq18}
\end{equation}

\noindent where $x$ and $\widetilde{x}$ denotes the ground truth image and the reconstruction, respectively, and $n$ is the number of pixels in the reconstructed image. A higher PSNR value indicates better reconstruction quality.

The SSIM, which lies in the range [0, 1], is used to measure the similarity between the ground-truth image and the reconstruction image,

\begin{equation}
    \text{SSIM}(\widetilde{x}, x)=\frac{1}{M} \sum_{j=1}^M \frac{\left(2 \widetilde{\mu}_j \mu_j+C_1\right)\left(2 \Sigma_j+C_2\right)}{\left(\widetilde{\mu}_j^2+\mu_j^2+C_1\right)\left(\widetilde{\sigma}_j^2+\sigma_j^2+C_2\right)},
    \label{eq19}
\end{equation}

\noindent where $\widetilde{\mu}_j$ and $\mu_j$ are the average pixel intensities, $\widetilde{\sigma}_j^2$ and $\sigma_j^2$ represent the variances, and $\Sigma_j$ is the the covariance of $\widetilde{x}, x$ at the $j$-th local window.  The constants 2$C_1=(K_{1}L)^2$ and $C_2=(K_{2}L)^2$ tend to be zero to avoid instability. Following \cite{wang_image_2004} and \cite{leuschner_quantitative_2021}, we choose $K_1 = 0.01$, $K_2 = 0.03$, $L=\max(x)-\min(x)$, and the window size is $7\times 7$. A higher SSIM value indicates better reconstruction quality.

\subsection*{Reconstruction Results}
\emph{AAPM challenge data}: We randomly select three full-dose CT images from AAPM challenge data to evaluate effectiveness of the propose method with the X-ray source intensity $I_i=[1e3, 1e4, 5e4]$. From Fig.~\ref{aapm1_reco}, Fig.~\ref{aapm2_reco} and Fig.~\ref{aapm3_reco}, we can observe that the quality of the FBP reconstruction images degraded significant as the X-ray source intensity decreased, resulting in amplified noise and artifacts distributed throughout the entire image. As a post-processing method, TV achieves higher quality images by post-processing the reconstructed images through FBP. Another post-processing and supervised method RED-CNN, can effectively remove noise and artifacts, but it tends to smooth out some tiny structures. Although DIP is unsupervised and takes random noise as input, it can effectively remove noise while producing images with higher resolution than RED-CNN. Comparing the reconstructed results by different methods, we can see that the proposed method achieves the best performance in terms of noise and artifacts attenuation and preservation of tiny structures.

To better illustrate the effectiveness of the proposed method, we further demonstrate the zoomed-in results corresponding to the red box in each ground truth. As shown in Fig.~\ref{aapm1_reco}, Fig.~\ref{aapm2_reco} and Fig.~\ref{aapm3_reco}, the reconstructed results by FBP and TV are contaminated by noise and artifacts. Although RED-CNN and DIP can suppress the noise, many valuable details are smoothed out. In comparison, the proposed method achieves better reconstruction accuracy than the competitive methods. It is worth noting that although the ground-truth images are norm-dose CT images, slight noise and artifacts still remain in them. Furthermore, the reconstructed results by the proposed method outperform the ground-truth images in terms of resolution, particularly with $I_{i}=1e4$ and $5e4$.           

\begin{table*}[]
\renewcommand{\arraystretch}{1.5}
\centering
\caption{Quantitative Results (PNSR/SSIM) of Different Algorithms.}
\label{tab1}
\begin{tabular}{ccccccc}
\hline \hline
Data                     & $I_i$  & FBP        & TV         & RED-CNN    & DIP        & Proposed            \\ \hline
\multirow{3}{*}{AAPM-1}  & $1e3$  & 10.92/0.09 & 14.51/0.19 & 23.54/0.64 & 29.83/0.81 & \textbf{32.37/0.82} \\
                         & $1e4$ & 10.92/0.09 & 21.19/0.37 & 25.39/0.74 & 30.22/0.82 & \textbf{35.99/0.92} \\
                         & $5e4$ & 10.92/0.09 & 26.25/0.54 & 24.23/0.69 & 30.36/0.82 & \textbf{37.85/0.94} \\ \cline{2-7} 
\multirow{3}{*}{AAPM-2}  & $1e3$  & 9.75/0.12  & 12.86/0.16 & 24.89/0.68 & 31.46/\textbf{0.85} & \textbf{31.56}/0.79 \\
                         & $1e4$ & 9.75/0.12  & 19.32/0.31 & 27.49/0.79 & 30.33/0.83 & \textbf{35.30/0.92} \\
                         & $5e4$ & 9.75/0.12  & 25.21/0.49 & 25.89/0.75 & 31.94/0.85 & \textbf{38.06/0.94} \\ \cline{2-7} 
\multirow{3}{*}{AAPM-3}  & $1e3$  & 10.89/0.13 & 13.30/0.17 & 25.71/0.73 & \textbf{35.71}/\textbf{0.90} & 35.47/\textbf{0.90} \\
                         & $1e4$ & 10.89/0.13 & 19.99/0.30 & 28.35/0.83 & 36.24/0.91 & \textbf{38.07/0.93} \\
                         & $5e4$ & 10.89/0.13 & 26.03/0.47 & 26.64/0.80 & 35.20/0.90 & \textbf{37.72/0.94} \\ \cline{2-7} 
\multirow{3}{*}{LoDoPab-CT} & $1e3$  & 10.65/0.09 & 15.72/0.22 & 24.33/0.65 & 32.76/0.88 & \textbf{33.25/0.89} \\
                         & $1e4$ & 10.65/0.09 & 24.26/0.45 & 26.12/0.73 & 33.14/0.88 & \textbf{35.84/0.94} \\
                         & $5e4$ & 10.65/0.09 & 29.24/0.65 & 24.95/0.70 & 32.82/0.87 & \textbf{36.39/0.95} \\ \hline \hline
\end{tabular}
\end{table*}

\emph{LoDoPab-CT data}: For the LoDoPab-CT data, the reconstruction results are shown in Fig.~\ref{lodopab_reco}. From Fig.~\ref{lodopab_reco}, we can observe that the performance of each reconstruction method is similar to their performance for the above AAPM challenge data reconstruction. The reconstructed results by FBP and TV suffer from noise and artifacts, although TV can suppress a lot of noise. The textures and edges in the reconstructed results by RED-CNN are smoothed out, whereas DIP can remove noise and preserve tiny structures more effectively. The proposed method achieves the best performance with regard to noise suppression and preservation of tiny structures. Furthermore, the reconstruction errors (Fig.~\ref{lodopab_errors}) further demonstrate that FBP method sacrifices a lot of useful information. TV and RED-CNN can effectively improve the reconstructed results by FPB; however, TV can not preserver edges well, and RED-CNN tends to smooth edges and textures. DIP has slighter residual errors in terms of edges and textures. Compared with the competitive methods, the proposed method significantly decreases the reconstruction errors by FBP and achieves minimal reconstruction errors. 

To quantitatively analyze the performance of our method, we calculate the the PSNR and SSIM values of the above reconstruction results, including the AAPM challenge data and the LoDoPab-CT data. As shown in Table.~\ref{tab1}, our method achieves the highest PSNR and SSIM among the five approaches, except for the reconstruction task of AAPM-2 with respect of the SSIM under $I_i=1e3$ and of AAPM-3 with respect of the PSNR under $I_i=1e3$. Specifically, the SSIM and PSNR of DIP are 0.06 and 0.24 dB higher than those of the proposed method.  

\begin{figure*}[]
\centering
\includegraphics[width=\textwidth]{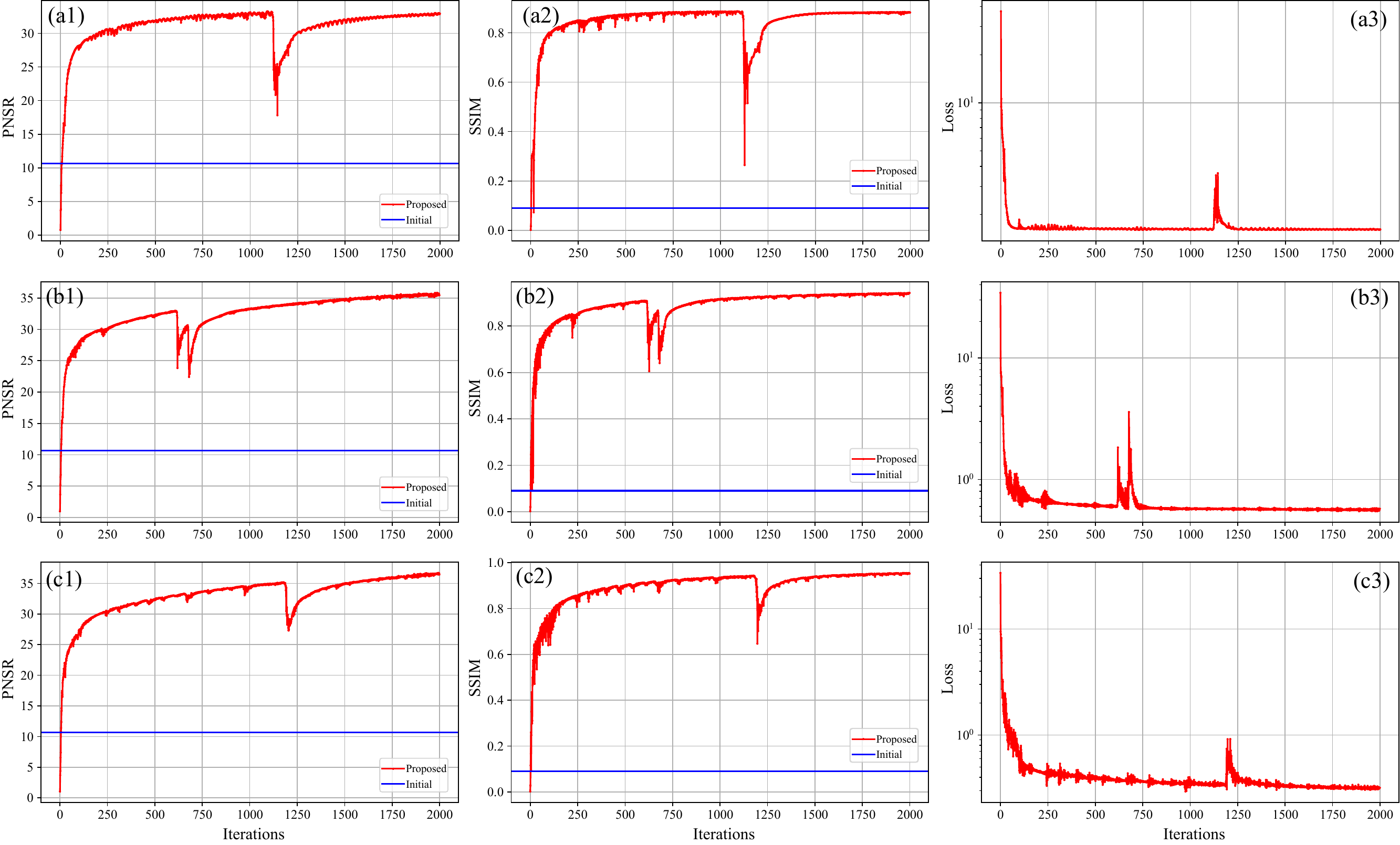}
\caption{Convergence analysis of the proposed method in LoDoPab-CT data reconstruction. (a1), (a2) and (a3) represent the PSNR, SSIM, and loss curve, respectively, for $I_i=1e3$. (b1), (b2) and (b3) depict the PSNR, SSIM, and loss curve, respectively, for $I_i=1e4$. (c1), (c2) and (c3) is the PSNR, SSIM, and loss curve, respectively, for $I_i=5e4$.}
\label{conver}
\end{figure*}

In addition, we take the evolution curves of PSNR and SSIM versus iteration of LoDoPab-CT data reconstruction as an example to illustrate the convergence of the proposed method. As shown in Fig.~\ref{conver}, the PSNR and SSIM increase while the loss decreases rapidly, which reveals that the proposed method can converge quickly. Specifically, the curves of PSNR and SSIM begin to converge after about 250 iterations, and the curves of loss start to converge after about 100 iterations. Although there are some fluctuations in these curves since the measurements contain noise, they converge quickly again, which indicates the good robustness of our method.  Table~\ref{tab2} lists the computation time of different method on a single GPU (Nvidia Tesla K80), it can be seen that FBP, TV and the proposed method have great disadvantages in terms of reconstruction time. Although RED-CNN only need one inference to reconstruct the high-quality image, the process of NN training is time consuming. Therefore, one can set a larger number of iterations to ensure that good reconstruction results can be obtained due to the rapid convergence and low computational cost of the proposed method.

\begin{table}[]
\renewcommand{\arraystretch}{1.5}
\centering
\caption{Computation Time of Different Algorithms for LoDoPab-CT Data Reconstruction.}
\label{tab2}
\setlength{\tabcolsep}{5.5mm}{\begin{tabular}{lcc}
\hline \hline
\multicolumn{1}{c}{\multirow{2}{*}{Method}} & \multicolumn{2}{c}{GPU time}   \\ \cline{2-3} 
\multicolumn{1}{c}{}                        & Training    & Reconstruction        \\ \hline
FBP (s)                                         & /           & 0.1              \\
TV (s/iteration)                                          & /           & 0.07  \\
RED-CNN (s/epoch)                                     & 310  & 5.1              \\
DIP (s/iteration)                                         & /           & 3.11  \\
Proposed (s/iteration)                                   & /           & 1.02  \\ \hline \hline
\end{tabular}}
\end{table}

\section*{Discussion and Conclusion}
For the initial LDCT reconstruction, we utilize the results reconstructed by FBP as the initial model for the proposed method. FBP can extract fundamental information about the internal structure of the human body, despite potential contamination from artifacts caused by the low intensity of X-ray. This is crucial for neural network-based LDCT imaging, as the black-box nature of these networks can significantly decrease the reliability of LDCT reconstruction results. It’s also important to note that the quality of the initial reconstructed image can impact the performance of the proposed method. One could substitute the FBP input with a high-quality image to further improve resolution. Additionally, FBP often performs much faster than iterative reconstruction approaches such as compressive sensing, which aids in enhancing inversion efficiency. Although our proposed method can converge rapidly, fluctuations due to noise in measurements might negatively impact the reconstruction efficiency. In future work, we aim to investigate better regularization techniques to promote convergence stability.

In this work, we propose an unsupervised and training data-free method for LDCT imaging. The proposed method aims to improve the initial reconstruction results with low quality, which reconstructs the high-quality image by DNN training without any training samples. We implement the DNN training by minimizing the $\ell_1$-norm distance between the CT measurements and their corresponding simulated sinogram data on the reconstructed image and the TV value of the reconstructed image. Notably, the proposed method dose not need to set weights for both the data fidelity term and the regularization term, which significantly reduces the difficulty of manually setting the weights. Experimental results on the AAPM challenge data and LoDoPab-CT data demonstrate that the proposed method could achieve better performance than the representative non-learning methods and supervised method, with higher resolution and lower computational cost. The proposed method can be implemented flexible and has the potential to be applied to other medical image reconstruction problems, including sparse-view CT reconstruction and image reconstruction from sparse samples in MRI. These applications are particularly useful when collecting training samples is either expensive or difficult.





\section*{Author Contributions}
Feng~Wang: Conceptualization, Investigation, Software, Validation, Methodology, Writing review \& editing.

\noindent Renfang~Wang: Conceptualization, Methodology, Supervision \& Writing review.

\noindent Hong~Qiu: Investigation, Methodology \& Writing review.


\bibliography{ref}

\begin{thebibliography}{10}

\bibitem{smith-bindman_radiation_2009}
Smith-Bindman R, Lipson J, Marcus R, Kim KP, Mahesh M, Gould R, et~al.
\newblock Radiation dose associated with common computed tomography examinations and the associated lifetime attributable risk of cancer.
\newblock Archives of Internal Medicine. 2009;169(22):2078--2086.
\newblock doi:{10.1001/archinternmed.2009.427}.

\bibitem{sidky_image_2008}
Sidky EY, Pan X.
\newblock Image reconstruction in circular cone-beam computed tomography by constrained, total-variation minimization.
\newblock Physics in Medicine and Biology. 2008;53(17):4777--4807.
\newblock doi:{10.1088/0031-9155/53/17/021}.

\bibitem{kim_sparse-view_2015}
Kim K, Ye JC, Worstell W, Ouyang J, Rakvongthai Y, Fakhri GE, et~al.
\newblock Sparse-{View} {Spectral} {CT} {Reconstruction} {Using} {Spectral} {Patch}-{Based} {Low}-{Rank} {Penalty}.
\newblock IEEE Transactions on Medical Imaging. 2015;34(3):748--760.
\newblock doi:{10.1109/TMI.2014.2380993}.

\bibitem{cai_cine_2014}
Cai JF, Jia X, Gao H, Jiang SB, Shen Z, Zhao H.
\newblock Cine {Cone} {Beam} {CT} {Reconstruction} {Using} {Low}-{Rank} {Matrix} {Factorization}: {Algorithm} and a {Proof}-of-{Principle} {Study}.
\newblock IEEE Transactions on Medical Imaging. 2014;33(8):1581--1591.
\newblock doi:{10.1109/TMI.2014.2319055}.

\bibitem{ma_iterative_2012}
Ma J, Zhang H, Gao Y, Huang J, Liang Z, Feng Q, et~al.
\newblock Iterative image reconstruction for cerebral perfusion {CT} using a pre-contrast scan induced edge-preserving prior.
\newblock Physics in Medicine and Biology. 2012;57(22):7519--7542.
\newblock doi:{10.1088/0031-9155/57/22/7519}.

\bibitem{zhang_spectral_2016}
Zhang Y, Xi Y, Yang Q, Cong W, Zhou J, Wang G.
\newblock Spectral {CT} {Reconstruction} {With} {Image} {Sparsity} and {Spectral} {Mean}.
\newblock IEEE Transactions on Computational Imaging. 2016;2(4):510--523.
\newblock doi:{10.1109/TCI.2016.2609414}.

\bibitem{qiong_xu_low-dose_2012}
{Qiong Xu}, {Hengyong Yu}, {Xuanqin Mou}, {Lei Zhang}, {Jiang Hsieh}, {Ge Wang}.
\newblock Low-{Dose} {X}-ray {CT} {Reconstruction} via {Dictionary} {Learning}.
\newblock IEEE Transactions on Medical Imaging. 2012;31(9):1682--1697.
\newblock doi:{10.1109/TMI.2012.2195669}.

\bibitem{yuan_low-dose_2019}
Yuan N, Zhou J, Gong K, Qi J.
\newblock Low-dose {CT} count-domain denoising via convolutional neural network with filter loss.
\newblock In: Bosmans H, Chen GH, Gilat~Schmidt T, editors. Medical {Imaging} 2019: {Physics} of {Medical} {Imaging}. San Diego, United States: SPIE; 2019. p.~26.
\newblock Available from: \url{https://www.spiedigitallibrary.org/conference-proceedings-of-spie/10948/2513479/Low-dose-CT-count-domain-denoising-via-convolutional-neural-network/10.1117/12.2513479.full}.

\bibitem{liu_sparse-sampling_2020}
Liu J, Li J.
\newblock Sparse-sampling {CT} {Sinogram} {Completion} using {Generative} {Adversarial} {Networks}.
\newblock In: 2020 13th {International} {Congress} on {Image} and {Signal} {Processing}, {BioMedical} {Engineering} and {Informatics} ({CISP}-{BMEI}). Chengdu, China: IEEE; 2020. p. 640--644.
\newblock Available from: \url{https://ieeexplore.ieee.org/document/9263571/}.

\bibitem{jin_deep_2017}
Jin KH, McCann MT, Froustey E, Unser M.
\newblock Deep {Convolutional} {Neural} {Network} for {Inverse} {Problems} in {Imaging}.
\newblock IEEE Transactions on Image Processing. 2017;26(9):4509--4522.
\newblock doi:{10.1109/TIP.2017.2713099}.

\bibitem{chen_low-dose_2017}
Chen H, Zhang Y, Kalra MK, Lin F, Chen Y, Liao P, et~al.
\newblock Low-{Dose} {CT} {With} a {Residual} {Encoder}-{Decoder} {Convolutional} {Neural} {Network}.
\newblock IEEE Transactions on Medical Imaging. 2017;36(12):2524--2535.
\newblock doi:{10.1109/TMI.2017.2715284}.

\bibitem{hasan_hybrid-collaborative_2021}
Hasan AM, Mohebbian MR, Wahid KA, Babyn P.
\newblock Hybrid-{Collaborative} {Noise2Noise} {Denoiser} for {Low}-{Dose} {CT} {Images}.
\newblock IEEE Transactions on Radiation and Plasma Medical Sciences. 2021;5(2):235--244.
\newblock doi:{10.1109/TRPMS.2020.3002178}.

\bibitem{lehtinen_noise2noise_2018}
Lehtinen J, Munkberg J, Hasselgren J, Laine S, Karras T, Aittala M, et~al.. {Noise2Noise}: {Learning} {Image} {Restoration} without {Clean} {Data}; 2018.
\newblock Available from: \url{http://arxiv.org/abs/1803.04189}.

\bibitem{hendriksen_noise2inverse_2020}
Hendriksen AA, Pelt DM, Batenburg KJ.
\newblock {Noise2Inverse}: {Self}-supervised deep convolutional denoising for tomography.
\newblock IEEE Transactions on Computational Imaging. 2020;6:1320--1335.
\newblock doi:{10.1109/TCI.2020.3019647}.

\bibitem{evangelista_rising_2023}
Evangelista D, Morotti E, Loli~Piccolomini E.
\newblock {RISING}: {A} new framework for model-based few-view {CT} image reconstruction with deep learning.
\newblock Computerized Medical Imaging and Graphics. 2023;103:102156.
\newblock doi:{10.1016/j.compmedimag.2022.102156}.

\bibitem{kang_deep_2017}
Kang E, Min J, Ye JC.
\newblock A deep convolutional neural network using directional wavelets for low-dose {X}-ray {CT} reconstruction.
\newblock Medical Physics. 2017;44(10):e360--e375.
\newblock doi:{10.1002/mp.12344}.

\bibitem{zheng_dual-domain_2020}
Zheng A, Gao H, Zhang L, Xing Y.
\newblock A dual-domain deep learning-based reconstruction method for fully {3D} sparse data helical {CT}.
\newblock Physics in Medicine and Biology. 2020;65(24):245030.
\newblock doi:{10.1088/1361-6560/ab8fc1}.

\bibitem{yuan_half2half_2020}
Yuan N, Zhou J, Qi J.
\newblock {Half2Half}: deep neural network based {CT} image denoising without independent reference data.
\newblock Physics in Medicine \& Biology. 2020;65(21):215020.
\newblock doi:{10.1088/1361-6560/aba939}.

\bibitem{zhu_image_2018}
Zhu B, Liu JZ, Cauley SF, Rosen BR, Rosen MS.
\newblock Image reconstruction by domain-transform manifold learning.
\newblock Nature. 2018;555(7697):487--492.
\newblock doi:{10.1038/nature25988}.

\bibitem{kandarpa_dug-recon_2021}
Kandarpa VSS, Bousse A, Benoit D, Visvikis D.
\newblock {DUG}-{RECON}: {A} {Framework} for {Direct} {Image} {Reconstruction} {Using} {Convolutional} {Generative} {Networks}.
\newblock IEEE Transactions on Radiation and Plasma Medical Sciences. 2021;5(1):44--53.
\newblock doi:{10.1109/TRPMS.2020.3033172}.

\bibitem{wu_computationally_2019}
Wu D, Kim K, Li Q.
\newblock Computationally efficient deep neural network for computed tomography image reconstruction.
\newblock Medical Physics. 2019;46(11):4763--4776.
\newblock doi:{10.1002/mp.13627}.

\bibitem{cheng_learned_2020}
Cheng W, Wang Y, Li H, Duan Y.
\newblock Learned {Full}-{Sampling} {Reconstruction} {From} {Incomplete} {Data}.
\newblock IEEE Transactions on Computational Imaging. 2020;6:945--957.
\newblock doi:{10.1109/TCI.2020.2996751}.

\bibitem{ye_unified_2021}
Ye S, Li Z, McCann MT, Long Y, Ravishankar S.
\newblock Unified {Supervised}-{Unsupervised} ({SUPER}) {Learning} for {X}-{Ray} {CT} {Image} {Reconstruction}.
\newblock IEEE Transactions on Medical Imaging. 2021;40(11):2986--3001.
\newblock doi:{10.1109/TMI.2021.3095310}.

\bibitem{ulyanov_deep_2020}
Ulyanov D, Vedaldi A, Lempitsky V.
\newblock Deep {Image} {Prior}.
\newblock International Journal of Computer Vision. 2020;128(7):1867--1888.
\newblock doi:{10.1007/s11263-020-01303-4}.

\bibitem{gong_pet_2019}
Gong K, Catana C, Qi J, Li Q.
\newblock {PET} {Image} {Reconstruction} {Using} {Deep} {Image} {Prior}.
\newblock IEEE Transactions on Medical Imaging. 2019;38(7):1655--1665.
\newblock doi:{10.1109/TMI.2018.2888491}.

\bibitem{yokota_dynamic_2019}
Yokota T, Kawai K, Sakata M, Kimura Y, Hontani H.
\newblock Dynamic {PET} {Image} {Reconstruction} {Using} {Nonnegative} {Matrix} {Factorization} {Incorporated} {With} {Deep} {Image} {Prior}.
\newblock In: 2019 {IEEE}/{CVF} {International} {Conference} on {Computer} {Vision} ({ICCV}); 2019. p. 3126--3135.

\bibitem{yoo_time-dependent_2021}
Yoo J, Jin KH, Gupta H, Yerly J, Stuber M, Unser M.
\newblock Time-{Dependent} {Deep} {Image} {Prior} for {Dynamic} {MRI}.
\newblock IEEE Transactions on Medical Imaging. 2021;40(12):3337--3348.
\newblock doi:{10.1109/TMI.2021.3084288}.

\bibitem{ding_dataset-free_2022}
Ding Q, Ji H, Quan Y, Zhang X.
\newblock A dataset-free deep learning method for low-dose {CT} image reconstruction.
\newblock Inverse Problems. 2022;38(10):104003.
\newblock doi:{10.1088/1361-6420/ac8ac6}.

\bibitem{leuschner_lodopab-ct_2021}
Leuschner J, Schmidt M, Baguer DO, Maass P.
\newblock {LoDoPaB}-{CT}, a benchmark dataset for low-dose computed tomography reconstruction.
\newblock Scientific Data. 2021;8(1):109.
\newblock doi:{10.1038/s41597-021-00893-z}.

\bibitem{baguer_computed_2020}
Baguer DO, Leuschner J, Schmidt M.
\newblock Computed tomography reconstruction using deep image prior and learned reconstruction methods.
\newblock Inverse Problems. 2020;36(9):094004.
\newblock doi:{10.1088/1361-6420/aba415}.

\bibitem{paszke_pytorch_2019}
Paszke A, Gross S, Massa F, Lerer A, Bradbury J, Chanan G, et~al.
\newblock {PyTorch}: {An} {Imperative} {Style}, {High}-{Performance} {Deep} {Learning} {Library}.
\newblock In: Advances in {Neural} {Information} {Processing} {Systems} 32. Curran Associates, Inc.; 2019. p. 8024--8035.
\newblock Available from: \url{http://papers.neurips.cc/paper/9015-pytorch-an-imperative-style-high-performance-deep-learning-library.pdf}.

\bibitem{he_iterative_2022}
He Z, Zhang Y, Guan Y, Guan B, Niu S, Zhang Y, et~al.
\newblock Iterative {Reconstruction} for {Low}-{Dose} {CT} {Using} {Deep} {Gradient} {Priors} of {Generative} {Model}.
\newblock IEEE Transactions on Radiation and Plasma Medical Sciences. 2022;6(7):741--754.
\newblock doi:{10.1109/TRPMS.2022.3148373}.

\bibitem{wang_image_2004}
Wang Z, Bovik AC, Sheikh HR, Simoncelli EP.
\newblock Image {Quality} {Assessment}: {From} {Error} {Visibility} to {Structural} {Similarity}.
\newblock IEEE Transactions on Image Processing. 2004;13(4):600--612.
\newblock doi:{10.1109/TIP.2003.819861}.

\bibitem{leuschner_quantitative_2021}
Leuschner J, Schmidt M, Ganguly PS, Andriiashen V, Coban SB, Denker A, et~al.
\newblock Quantitative {Comparison} of {Deep} {Learning}-{Based} {Image} {Reconstruction} {Methods} for {Low}-{Dose} and {Sparse}-{Angle} {CT} {Applications}.
\newblock Journal of Imaging. 2021;7(3):44.
\newblock doi:{10.3390/jimaging7030044}.

\end{thebibliography}

\end{document}